\documentclass{article}
\usepackage{graphicx}
\usepackage{authblk}
\usepackage{xspace}
\title{PyHEP.dev 2024 Workshop Summary Report, August 26--30 2024, Aachen, Germany}
\date{September 2024}

\def\abcdpyhf   {\mbox{\texttt{abcd\_pyhf}}\xspace}
\def\Airflow   {\mbox{\texttt{Airflow}}\xspace}
\def\AwkwardArray   {\mbox{\texttt{Awkward Array}}\xspace}
\def\BLUE   {\mbox{\texttt{BLUE}}\xspace}
\def\boosthistogram {\mbox{\texttt{boost-histogram}}\xspace}
\def\cabinetry   {\mbox{\texttt{cabinetry}}\xspace}
\def\Coffea   {\mbox{\texttt{Coffea}}\xspace}
\def\columnflow   {\mbox{\texttt{columnflow}}\xspace}
\def\cudahistogram   {\mbox{\texttt{cuda-histogram}}\xspace}
\def\Dask   {\mbox{\texttt{Dask}}\xspace}
\def\daskawkward   {\mbox{\texttt{dask-awkward}}\xspace}
\def\daskhistogram   {\mbox{\texttt{dask-histogram}}\xspace}
\def\dill   {\mbox{\texttt{dill}}\xspace}
\def\evermore   {\mbox{\texttt{evermore}}\xspace}
\def\Gammapy   {\mbox{\texttt{Gammapy}}\xspace}
\def\hfivepy   {\mbox{\texttt{h5py}}\xspace}
\def\Hist   {\mbox{\texttt{Hist}}\xspace}
\def\HistFitter   {\mbox{\texttt{HistFitter}}\xspace}
\def\LAW   {\mbox{\texttt{LAW}}\xspace}
\def\Luigi   {\mbox{\texttt{Luigi}}\xspace}
\def\Btoluigi   {\mbox{\texttt{b2luigi}}\xspace}
\def\JAX   {\mbox{\texttt{JAX}}\xspace}
\def\mplhep   {\mbox{\texttt{mplhep}}\xspace}

\def\PocketCoffea   {\mbox{\texttt{PocketCoffea}}\xspace}
\def\pyhf   {\mbox{\texttt{pyhf}}\xspace}
\def\Ray {\mbox{\texttt{Ray}}\xspace}
\def\RDataFrame {\mbox{\texttt{RDataFrame}}\xspace}
\def\RooFit {\mbox{\texttt{RooFit}}\xspace}
\def\Snakemake   {\mbox{\texttt{Snakemake}}\xspace}
\def\Uproot   {\mbox{\texttt{Uproot}}\xspace}
\def\Vector   {\mbox{\texttt{Vector}}\xspace}
\def\TaskVine   {\mbox{\texttt{TaskVine}}\xspace}
\def\TensorFlow   {\mbox{\texttt{TensorFlow}}\xspace}

\def\xRooFit   {\mbox{\texttt{xRooFit}}\xspace}
\def\Zarr   {\mbox{\texttt{Zarr}}\xspace}
\def\zfit   {\mbox{\texttt{zfit}}\xspace}
\def\JSON   {\mbox{\texttt{JSON}}\xspace}
\def\YAML   {\mbox{\texttt{YAML}}\xspace}
\def\NumFOCUS   {\mbox{NumFOCUS}\xspace}
\def\Quansight   {\mbox{Quansight}\xspace}
\def\QuantStack   {\mbox{QuantStack}\xspace}
\def\Anaconda   {\mbox{Anaconda}\xspace}
\def\Kubernetes   {\mbox{\texttt{Kubernetes}}\xspace}
\def\CoffeaCasa   {\mbox{\texttt{coffea-casa}}\xspace}
\def\CMSCombine   {\mbox{\texttt{CMS Combine}}\xspace}
\def\Combiner   {\mbox{\texttt{Combiner}}\xspace}
\def\RooUnfold   {\mbox{\texttt{RooUnfold}}\xspace}
\def\HAverage   {\mbox{\texttt{HAverager}}\xspace}
\def\HTCondor   {\mbox{\texttt{HTCondor}}\xspace}
\def\LSF   {\mbox{\texttt{LSF}}\xspace}

\def\Slurm   {\mbox{\texttt{Slurm}}\xspace}
\def\ARC   {\mbox{\texttt{ARC}}\xspace}
\def\gLite   {\mbox{\texttt{gLite}}\xspace}
\def\CMSCRAB   {\mbox{\texttt{CMS-CRAB}}\xspace}
\def\WLCG   {\mbox{\texttt{WLCG}}\xspace}
\def\ROOT       {\mbox{\texttt{ROOT}}\xspace}
\def\HDFf {\mbox{\texttt{HDF5}}\xspace}
\def\plothist {\mbox{\texttt{plothist}}\xspace}
\def\Matplotlib {\mbox{\texttt{Matplotlib}}\xspace}
\def\Python {\mbox{Python}\xspace}
\def\arrays {\mbox{\texttt{arrays}}\xspace}
\def\Blosctwo {\mbox{\texttt{Blosc2}}\xspace}
\def\ApacheSpark {\mbox{Apache Spark}\xspace}
\def\StellarHPX {\mbox{\texttt{Stellar HPX}}\xspace}
\def\Spark {\mbox{Spark}\xspace}
\def\Jupyter {\mbox{Jupyter}\xspace}
\def\Julia {\mbox{Julia}\xspace}
\def\cpp {\mbox{C\texttt{++}}\xspace}
\def\Yadage {\mbox{\texttt{Yadage}}\xspace}
\def\RECAST {\mbox{\texttt{RECAST}}\xspace}

\def\lhc    {\mbox{LHC}\xspace}
\def\lhcb {\mbox{LHCb}\xspace}
\def\atlas  {\mbox{ATLAS}\xspace}
\def\cms    {\mbox{CMS}\xspace}

\def\belletwo  {\mbox{Belle II}\xspace}

\DeclareUnicodeCharacter{0301}{*************************************}
\usepackage{jheppub}
\usepackage{hyperref}
\usepackage{graphicx}
\usepackage[titletoc]{appendix}
\usepackage{xspace}
\usepackage{enumitem}
\setlist{noitemsep}
\usepackage[T1]{fontenc} 
\usepackage[style=numeric,sorting=none]{biblatex}
\begin{document}

\author[1]{\small Azzah Alshehri}
\author[2]{\small Jan Bürger}
\author[3]{\small Saransh Chopra}
\author[4]{\small Niclas Eich}
\author[5]{\small Jonas Eppelt}
\author[6]{\small Martin Erdmann}
\author[7]{\small Jonas Eschle}
\author[2, 4]{\small Peter Fackeldey}
\author[4]{\small Maté Farkas}
\author[8]{\small Matthew Feickert}
\author[9]{\small Tristan Fillinger}
\author[4]{\small Benjamin Fischer}
\author[2, 10]{\small Stefan Fr\"ose}
\author[11]{\small Lino Oscar Gerlach}
\author[12]{\small Nikolai Hartmann}
\author[5]{\small Alexander Heidelbach}
\author[8]{\small Alexander Held}
\author[13]{\small Marian I Ivanov}
\author[14]{\small Josué Molina}
\author[4]{\small Yaroslav Nikitenko}
\author[11]{\small Ianna Osborne}
\author[15]{\small Vincenzo Eduardo Padulano}
\author[11]{\small Jim Pivarski}
\author[9]{\small Cyrille Praz}
\author[16]{\small Marcel Rieger}
\author[17]{\small Eduardo Rodrigues}
\author[18]{\small Oksana Shadura}
\author[15]{\small Juraj Smieško}
\author[19]{\small Giordon Holtsberg Stark}
\author[2]{\small Judith Steinfeld}
\author[2]{\small Angela Warkentin}

\affiliation[1]{\footnotesize University of Hafr AlBatin}
\affiliation[2]{\footnotesize ErUM-Data-Hub}
\affiliation[3]{\footnotesize University College London (UCL)}
\affiliation[4]{\footnotesize RWTH Aachen University}
\affiliation[5]{\footnotesize Karlsruher Insititute of Technology (KIT)}
\affiliation[6]{\footnotesize Rheinisch Westfaelische Tech. Hoch.}
\affiliation[7]{\footnotesize Syracuse University}
\affiliation[8]{\footnotesize University of Wisconsin-Madison}
\affiliation[9]{\footnotesize High Energy Accelerator Research Organization (KEK)}
\affiliation[10]{\footnotesize TU Dortmund University}
\affiliation[11]{\footnotesize Princeton University}
\affiliation[12]{\footnotesize Ludwig Maximilians Universitat}
\affiliation[13]{\footnotesize GSI - Helmholtzzentrum fur Schwerionenforschung GmbH}
\affiliation[14]{\footnotesize EAP Zamorano}
\affiliation[15]{\footnotesize CERN}
\affiliation[16]{\footnotesize Hamburg University}
\affiliation[17]{\footnotesize Oliver Lodge Laboratory, University of Liverpool}
\affiliation[18]{\footnotesize University of Nebraska Lincoln}
\affiliation[19]{\footnotesize SCIPP, University of California Santa Cruz}

\abstract{The second PyHEP.dev workshop, part of the ``Python in HEP Developers'' series organized by the HEP Software Foundation (HSF), took place in Aachen, Germany, from August 26 to 30, 2024. This gathering brought together nearly 30 \Python package developers, maintainers, and power users to engage in informal discussions about current trends in Python, with a primary focus on analysis tools and techniques in High Energy Physics (HEP).

The workshop agenda encompassed a range of topics, such as defining the scope of HEP data analysis, exploring the Analysis Grand Challenge project, evaluating statistical models and serialization methods, assessing workflow management systems, examining histogramming practices, and investigating distributed processing tools like \RDataFrame, \Coffea, and \Dask. Additionally, the workshop dedicated time to brainstorming the organization of future PyHEP.dev events, upholding the tradition of alternating between Europe and the United States as host locations.

This document, prepared by the session conveners in the weeks following the workshop, serves as a summary of the key discussions, salient points, and conclusions that emerged.}

\maketitle

\newpage
\section{Introduction}

The PyHEP.dev workshop~\cite{PyHEP2024dev}, the second in the ``Python in HEP developers'' workshop series of the HEP Software Foundation (HSF)~\cite{hsf}, took place in Aachen, Germany, on August 26--30, 2024. Thirty \Python package developers, maintainers, and power users, all of whom are authors of this report, gathered together in an informal setting to discuss relevant and timely trends in \Python, largely targeting end-user analysis. The following topics were central points of discussion: how PyHEP does (or does not) relate to physicists' needs; the Analysis Grand Challenges that run analyses at scale at analysis facilities; leveraging key packages from the ecosystem; the development of statistical packages, models, interfaces, and serialization; workflow management systems; histogramming; and key distributed processing tools like \RDataFrame~\cite{rdataframe}, \Coffea~\cite{coffea}, and \Dask~\cite{dask}. Finally, we brainstormed the organization of future PyHEP.dev workshops and how to connect PyHEP activities with the larger scientific software developer community.

Participation was in-person, but a Zoom connection with an \textit{Owl} camera also broadcast all sessions to participants who could not attend. About four to five people joined remotely, mostly listening, though one talk was presented from Japan.

In the week following the workshop, the organizers curated this document, which is meant to be a summary reflecting the main discussions, the key points raised, and our collective conclusions. The document was written collaboratively by nearly all participants, refined into a coherent whole, and reviewed by all participants in its final form before it was made public.

\section{Scope of the workshop}

Our first discussion on Monday aimed to define the scope of the workshop, ensuring that all participants shared a common understanding of what ``HEP data analysis'' entails, both in terms of software and infrastructure. For the purpose of the workshop, we focused on the following understanding of the term:


\begin{quote} Data analysis transforms raw measurements into human-interpretable formats, such as tables and visualizations, for the extraction of physics quantities of interest. The process includes summary statistics, statistical inference, and machine learning. The act of analyzing data is highly iterative, changing strategy in response to partial results. \end{quote}

\noindent We interpreted High Energy Physics (HEP) broadly to include any experimental physics of fundamental particles. Many --- but not all --- of the PyHEP.dev workshop participants collaborate on \lhc (Large Hadron Collider~\cite{lhc}) and flavor physics experiments, primarily \atlas~\cite{atlas}, \cms~\cite{cms}, \lhcb~\cite{lhcb}, and \belletwo~\cite{belle2}. In particular, the tools developed by PyHEP.dev participants target the phase of work after central production (e.g.,\ after triggers and event reconstruction) through final publication.

During the discussion, we recognized that participants had different views about whether 
high-level physics/statistics should be strictly separated from low-level computing details.
Some 5–10 years ago, there was a much stronger consensus that data analysts should be completely isolated from computing details, possibly through a new domain-specific language specialized for HEP. In this model, everything concerning where and how data are loaded and computations are performed would be handled automatically. We noted that this is no longer a goal shared by all developers; some physicist-users will need to be aware of certain computational details. Specifically, it needs to be possible to break through the high-level abstraction to control low-level details in some situations, particularly to optimize for best performance and to enable debugging. This was crucial for the ``200 Gbps Analysis Challenge'' run by the Institute for Research and Innovation in Software for High Energy Physics (IRIS-HEP) in Spring 2024~\cite{AGC200Gbps}, which required significant debugging and coordination between the analyzers and the system facilitators (acknowledging that it was the first test of its kind). We did agree, though, that HEP analysis tools should initially or primarily present a high-level view, at least for first-time users, in which low-level computing details are only encountered by choice.

It was mentioned that visualizations do not need to be limited to two-dimensional plots. It was also proposed to improve physics visualizations by using graphics formats native to final publications, such as PGF/TikZ~\cite{pgftikz} for \LaTeX~\cite{latex} ~articles (publications and Beamer~\cite{beamer} slides).

Another aspect of HEP analysis is its long-term nature. Data safety --- that is, synchronization and backups --- was also mentioned in a dedicated presentation as an essential condition for analysis preservation.

Due to the collaborative nature of physics data analysis, data exchange formats were discussed at length. Agreement on common formats, and easy conversions between them, are key to interoperability. In particular, we talked about the HS3 HEP Statistics Serialization Standard~\cite{HS3} in \JSON/\YAML, \Dask graphs for delayed computations, and \AwkwardArray\!\!s~\cite{awkwardarray} for \Python-\Julia integration. Although some users and developers use \Python's pickle format, especially for temporary storage or transmitting data over a network, pickle is not stable across software version updates. It is not interchangeable with \cpp and other languages, and it is fundamentally unsafe, since loading a pickle file can execute arbitrary code. A similar complaint was raised about \YAML, which can run arbitrary code, but most \YAML deserializers have a \texttt{safe\_load} function, which refuses to run code embedded in the \YAML document. Developers should use such functions exclusively. (The equivalent option does not exist for pickle.)

We also discussed the transition of software from development stages to maintenance stages of its life cycle. Some \Python projects that began in an exploratory way had to stabilize to remain useful to those who adopted them, either direct users or downstream dependencies; interface changes would disrupt these users. Maintaining stable software is a different activity from developing new, prospective software, requiring a different class of developers, such as postdoctoral researchers and Research Software Engineers (RSEs), rather than undergraduate or graduate students. In many cases, it requires different funding, since most funding sources target research and development. The problem is not unique to HEP; it also affects Astronomy and the Astropy project~\cite{Astropy} in particular. While there are some organizations that sponsor or fund open-source scientific \Python projects --- notably \NumFOCUS~\cite{NumFOCUS}, \Quansight~\cite{Quansight}, \Anaconda~\cite{Anaconda}, and \QuantStack~\cite{QuantStack} --- none of these are known to fund software developers outside of their organizations. \NumFOCUS is a non-profit organization that helps with financial administration and legal services, while \Quansight, \Anaconda, and \QuantStack are commercial companies that only employ a few open-source developers; most of their revenue comes from consulting. At the end of this discussion, we did not identify, unfortunately, any funding sources that could support postdoctoral researchers and RSEs to maintain foundational \Python software for analysis.

Finally, we discussed the impact that large computations have on climate change~\cite{climate}. We cannot ignore the environmental impact of our research, especially the significant CO$_2$ emissions and energy consumption associated with the powerful computing resources we rely on. Recognizing that our actions have consequences, we need to act decisively and quickly to mitigate the damage we are causing. We need to raise awareness, share knowledge, and optimize our data and code production in the short term. Looking further ahead, we must adapt our computing practices to align with the availability of renewable energy sources. We concluded that the time to act is now, and we must work together to ensure a sustainable future for both our research and our planet.

\section{Analysis Grand Challenge}

HEP analyses, as described in the previous section, vary widely depending on the target experiment, the university group, or specific user requirements. This makes it difficult to quantify the properties of ``a typical HEP analysis'', though some kind of benchmark is needed to guide discussion about what is easy or hard for physicists to do, what has acceptable performance, and what needs improvement. In this regard, the Analysis Grand Challenge (AGC) project~\cite{AGC} defines a sample analysis task that captures aspects of relevant physics analysis problems, including the treatment of systematic variations, statistical modelling and inference, and machine learning. A variety of implementations have been developed for this task, allowing us to probe user experience and interoperability, and helping to center community discussions around a common benchmark. The first implementation makes use of many tools in the \Python HEP ecosystem and, in particular, a stack of Scikit-HEP libraries~\cite{scikit-hep,Rodrigues:2020syo}. Although the workflow shown in the AGC is mostly based around \atlas and \cms physics analyses, it still serves as a tutorial that shows how to build an end-to-end HEP analysis pipeline more generally, since it stresses all of the computational features of a HEP analysis task~\cite{AGCwishlist}.

Both the AGC presentation and the following demo brought up many questions from the PyHEP.dev participants, some of whom were introduced to \Dask for the first time. The demo of the AGC implementation using \daskawkward~\cite{daskawkward} included simple examples of that library~\cite{IDAP}. It highlighted how \Uproot~\cite{Uproot}, \AwkwardArray, \Dask, and \Coffea can be used together for data processing and visualization. Upon seeing the \Dask visualizations, some participants were concerned that users would have to become experts in a whole new set of monitoring tools, but we concluded that \Dask visualizations are for workflow debugging and optimization, not for physics. Key points were raised about the transition from vectors implemented in \Coffea to the \Vector~\cite{Vector} library and the need for static type-checking in \AwkwardArray's \texttt{behaviors}.

The discussion about distributed computing and the approach used for the IRIS-HEP Data Analysis Pipeline (IDAP) 200 Gbps Challenge~\cite{IDAP2} spanned a broad range of topics: caching strategies, handling the memory footprint, and the Aachen team's approach of using high-memory \Dask workers for giant histograms.

The AGC project revealed a number of topics to consider when running analysis pipelines in this format across a variety of facilities, including the following:
\begin{itemize}
    \item \textbf{memory footprint}, which relates to choosing a suitable number of partitions for each dataset;
    \item \textbf{integrating heterogeneous computing resources}, such as GPUs;
    \item \textbf{scaling strategies}, such as automatic scaling provided by \Dask.
\end{itemize}

The best configuration may vary by facility and use case. Interactions with companies like Coiled.io~\cite{Coiled}, the core \Dask team or the broader \Dask community may also yield useful insights for good patterns to adopt.

A discussion  followed on  differences between \Dask's collections (\texttt{dask.array}, \texttt{dask.frame}, \texttt{dask.bag}, and \texttt{dask\_awkward}) and how \Dask handles partitions. We discussed performance profiling, focusing on bottlenecks such as decompression and array concatenation. The scalability of the \Dask network, especially with \Kubernetes, was also examined. Concerns about \Dask's memory limits for workers and the potential for heterogeneity were raised, with suggestions to get support from \Dask experts beyond HEP. The session emphasized the importance of addressing these challenges and keeping track of them for future collaboration and problem-solving.

The PyHEP.dev participants were interested in having more transparent environment handling for easier debugging and development at \CoffeaCasa~\cite{coffeacasa} analysis facilities.

\section{Statistical Models, Interfaces and Serialization}

The HEP community uses many different statistical tools. At PyHEP.dev we discussed some of the most popular packages and how they interact with each other. In the following, we present how these tools can fit together to build statistical models and perform experiment-specific inference. In addition, we also discuss how models may be serialized for out-of-core computations and later retrieval.

First, we made a distinction between ``open-world'' probability tools and specialized ones. Specialized tools target a specific statistical problem, and the limited scope allows them to reduce boilerplate code or even be configured declaratively, which eases the path to serialization. Open-world tools are intended for users to construct their own specialized models, and that open-ended scope makes it harder to encapsulate all the things the tool can do, especially if it allows users to write their own functions. \RooFit~\cite{Verkerke:2003ir} and \zfit~\cite{zfit} are examples of open-world tools, as they allow users to build a likelihood using any probability distribution, both using pre-defined components in the library as well as arbitrary functions. Thus, they allow both binned and unbinned statistical inference. However, they necessarily have a steeper learning curve, which can be mitigated by higher-level interface layers such as \xRooFit~\cite{xRooFit}. New cross-cutting features are harder to add to open-world tools since they must ensure support across the breadth of the library.

This led to the development of specialized tools that intentionally constrain their scope for specific purposes, such as \pyhf~\cite{pyhf_joss}, \cabinetry~\cite{cabinetry}, \evermore~\cite{evermore}, \Gammapy~\cite{Gammapy}, \CMSCombine~\cite{CMS:2024onh}, \HistFitter~\cite{histfit}, \Combiner~\cite{combiner}, \RooUnfold~\cite{roounfold}, \HAverage~\cite{haverage,haverager_paper}, and \BLUE~\cite{blue}. The scopes of these tools partially overlap. For example, while a specialized tool like \pyhf is flexible enough to allow data-driven estimation of fake events via the ABCD method~\cite{abcd}, the interface is challenging for a typical user and a wrapper, \abcdpyhf~\cite{abcdpyhf}, was written to smooth that experience.

Next, we considered function minimizers and statistical inference tools. There is a natural boundary between model-building, function minimizing, and statistical inference: building models is one well-defined scope, minimizing negative log-likelihoods is another, and interpreting the minimum as a statistical inference is yet another. Software can be better scoped to focus on each of these topics individually, but only if we can reach an agreement on protocols for common formats for likelihoods and their minima. By way of analogy, a similar factorization was possible in Scikit-HEP's histogram packages, which focus on fast histogram filling, plotting, slicing, projecting, and rebinning, thanks to agreements on the Unified Histogram Interface (UHI)~\cite{UHI}. An exchange format for likelihoods would have to standardize the call signature, parameter configurations (bounds, initial values, fixed parameters), and the information returned about the minimum of the likelihood. Statistical inference packages need to have enough information in this output to be specialized for particular experiments.

Although protocols only standardize \Python-callable interfaces, serialization formats specify byte-formats for analysis steps that can be saved for later processing, transmitted through a network, or passed between programs written in different languages. We discussed the ongoing effort to serialize statistical models using HS3 in \JSON/\YAML, but also considered the possibility that users would want to serialize hand-written functions. Formats such as StableHLO IR and LLVM IR encode functions that can be compiled and used in any language, by feeding them into industry-standard compiler chains --- XLA~\cite{xla} and LLVM~\cite{llvm}, respectively --- without any dependence on big libraries such as \JAX~\cite{jaxgithub,JAX} or \TensorFlow~\cite{tensorflow}. The \dill library~\cite{dill} can save executable \Python functions, but only as black boxes that can't be inspected or generalized. Also, whereas Intermediate Representations (IRs) can be restricted to mathematical computations only, functions saved with \dill can run arbitrary code. HS3 without arbitrary functions is easier to inspect and identify the physics-motivated parts than any serialization format for arbitrary functions, and providing both necessarily introduces two ways of doing the same thing. StableHLO IR guarantees backward compatibility.

The HS3 format is still under development and needs further consolidation. As of yet, serializing arbitrary IR functions is not part of the HS3 scope.

\section{Workflow Management Systems}

High-level physics analyses usually consist of a considerable amount of logically separated workloads. In general, the interface between these workloads does not rely on an event-by-event (chunk) data flow such as utilized by, for example, experiment software or parallel computing libraries like \Dask. The workloads to perform a specific analysis often form a loose collection of inhomogeneous procedures, encoded in executable files such as Shell and Python scripts, and are often executed manually. Hereby, their execution order is dictated by the dependencies between them in terms of the existence of results of one or more previous procedures. Beyond a certain scale and complexity, the manual steering of analysis workflows can be time-consuming, prone to errors, and potentially leads to undocumented relations between workloads.

A Workflow Management System (WMS)\footnote{Commonly referred to as ``workflow engines'' or ``workflow languages''.} is software to describe, manage, and execute a sequence of arbitrary workloads. It is agnostic to the concrete type of operations and the data these workloads are performed on. In the HEP context, a WMS allows one to describe all necessary steps of an analysis bridging Monte Carlo simulation (MC) production, reconstruction, and statistical analysis. Common implementations also provide tools to split execution logic from computing infrastructure.

\Dask can be understood as a specialized  WMS with more focus on the parallelization of loops and array-based calculation as well as the scalability of these processes. Consequently, general WMS engines and \Dask can be used complementary in HEP analysis workflows. This allows for both bridging heterogeneous workflows and profiting from \Dask's parallelization mechanisms.

Currently, several different WMS engines are used in HEP. At PyHEP.dev, \Luigi~\cite{luigi} was most thoroughly discussed (not counting \Dask), followed by \Snakemake~\cite{Snakemake,Snakemake_paper}, and Apache \Airflow~\cite{airflow} was only briefly mentioned. Users and maintainers of \Luigi extensions were present and talked about their work.

WMS engines are increasingly common in HEP, both for end-user analysis and for services such as calibration. We identified the following cases:
\begin{itemize}
    \item \textbf{\atlas}
    \begin{itemize}
        \item \Yadage~\cite{Cranmer:2017frf} workflows used in \RECAST~\cite{Cranmer:2010hk, RECAST_software} full-analysis pipelines, which most ATLAS searches for physics beyond the Standard Model implement as part of the analysis preservation process prior to publication.
    \end{itemize}
    \item \textbf{\cms}
    \begin{itemize}
        \item \LAW~\cite{LAW} workflows in end-user analysis, such as \columnflow~\cite{columnflow} and \PocketCoffea~\cite{PocketCoffea}.
    \end{itemize}
    \item \textbf{\lhcb}
    \begin{itemize}
        \item \Snakemake workflows in end-user analysis. In common cases, \Snakemake runs in GitLab CI, and one physics working group went as far as having a series of workflows implemented for many of the working group analyses, providing group-level maintenance of templates.
    \end{itemize}
    \item \textbf{\belletwo}
    \begin{itemize}
        \item \Btoluigi~\cite{b2luigi,b2luigi_paper} in automated, collaboration-wide processes, such as systematic uncertainty calculations and data-MC validation.
        \item \Btoluigi and \Snakemake in end-user analyses.
        \item Apache \Airflow steers the calibration framework.
    \end{itemize}
\end{itemize}

\Luigi-based workflows are primarily limited by the built-in scheduler's capacity to scale, up to about 10,000 dependencies. These workflows can't be further granularized, which puts an upper limit on the scope of debuggable code. Since no alternate schedulers exist for \Luigi-based workflows, scaling beyond this limit means rewriting the workflow. An example case where this comes up is per-dataset debugging. However, there are simple ways to overcome this limitation in practical cases such as step-wise execution of parts of exceedingly large dependency trees.

At this workshop, two WMS extensions were presented in detail: \Btoluigi and \LAW.

\Btoluigi extends \Luigi to provide specialized solutions for situations that arise in HEP. \Btoluigi integrates with \HTCondor~\cite{htcondor} and \LSF~\cite{lsf}, which are often found in HEP computing clusters, and it, therefore, allows users to write a workflow for both of these batch systems. Furthermore, the package provides mechanisms for task-specific and global settings, such as I/O, software environment, and batch system user configurations and quality-of-life improvements for analysts. \Btoluigi provides specialized interfaces for \belletwo workflows, allowing particularly smooth interaction with the \belletwo software framework. Workflows such as the automated validation framework, the systematic uncertainty framework, and many \belletwo physics analyses have been automated using \Btoluigi. Maintenance of \Btoluigi is now in the hands of the \belletwo collaboration and will be followed up by its software group.

\LAW stands for \Luigi Analysis Workflow, and it is also based on \Luigi. \LAW establishes a generic design pattern for analyses of arbitrary scale and complexity by providing building blocks to integrate with remote resources in a generic, interchangeable way so that it does not depend on a particular cluster infrastructure. In particular, it completely separates analysis algorithms from run locations, storage locations, and software environments. To cope with sophisticated demands of end-to-end HEP analyses, \LAW supports job execution on \WLCG infrastructure (\ARC, \gLite, \CMSCRAB)~\cite{wlcg} as well as local computing clusters (\HTCondor, \Slurm~\cite{slurm}, \LSF), remote file access via various protocols using the Grid File Access Library (GFAL2)~\cite{gfal2}, and an environment sandboxing mechanism with support for sub-shells and virtual environments, as well as Docker~\cite{docker} and Singularity~\cite{singularity} containers. \LAW ultimately aims to provide analysis preservation out-of-the-box. \LAW is developed in an open-source, experiment-agnostic way, and its user base has steadily increased over the past eight years.

\Btoluigi and \LAW implement similar functionalities differently, with \LAW presenting a more extensive feature set. While \Btoluigi intends to be a drop-in replacement for \Luigi, \LAW is more focused on extending the overall library with user-targeted features. What they have in common is extensive support for typical HEP batch systems, easier and more automated handling of \Luigi targets and steering of user environments.

We discussed a possible merge of these packages, though it would be quite challenging at this stage. Beyond the technological differences and differences in project goals, both packages already have many users who depend on them as foundational libraries.

We also discussed the future of these WMS extensions: parameter simplification, developments related to Python's Global Interpreter Lock (GIL), and upstreaming features to \Luigi.

On the topic of parameter simplification, the fact that large dependency trees generate a large number of parameters to configure complicates user code and makes workflows less flexible. We discussed ways to implicitly control the branching of Directed Acyclic Graphs (DAGs), expand the existing \texttt{Parameters} object in \Luigi, and describe complex dependency trees with aggregation schemes. We identified options to implement these ideas and planned to follow up on them together.

With ongoing developments in Python to make the GIL optional—such as PEP 703~\cite{pep703}—there is potential to improve the performance of core elements of \Luigi, such as the parallel determination of the dependency tree that constitutes the workflow. Currently, this computation is being carried out in a single thread, since the GIL would prevent performance advantages from multiple threads. (Multiprocessing is not a viable option because of the need to share data between threads.) Our community should identify whether the option to remove the GIL and increase parallel processing can improve the performance of \Luigi's DAG-building process.

\Btoluigi and \LAW extend \Luigi with features that, arguably, should be in \Luigi itself. We identified a handful of features that could be upstreamed. However, if the community decides to use \Luigi to a wider extent in mission-critical projects, we'll need to have a larger discussion with future stakeholders to unify requests for features for the HEP community with the rest of the world. Ideally, members of the HEP community would have to become core developers and maintainers of the \Luigi project.

\section{Histogramming}

Histogramming concerns three major topics: the production, presentation, and serialization of histograms. Additionally, the accumulation --- that is, merging --- of histograms from concurrent histogram--filling processes, where each process fills a separate copy of the histogram, into a single summed histogram set can be considered separately. All of these topics were raised during the presentations and the discussions that followed.

The production of histograms had once been a contentious topic in \Python, with dozens of libraries providing mutually-incompatible histogram objects. This changed with \texttt{Boost::Histogram} and its \Python interfaces, \boosthistogram~\cite{boost_histogram} and \Hist~\cite{Hist}. In addition, the Unified Histogram Interface (UHI) defined protocols to share boost-style histogram interfaces between libraries, so libraries that adhere to these protocols can be compatible without sharing memory layouts. For example, \daskhistogram implements delayed histograms and \cudahistogram implements histograms on GPUs, which are not \texttt{Boost::Histogram} objects, but can be readily converted. Currently, \texttt{Boost}-style histograms are still incompatible with \ROOT's~\cite{root} histograms (\Uproot's conversion from \texttt{Boost} to \ROOT is lossy), but \ROOT will be adopting UHI in the near future.~\cite{ROOT_UHI_2024}

\texttt{Boost-histogram} is now widely used for filling histograms in \Python on CPUs. Histogram-filling on GPUs with \cudahistogram is in development. Some cases, such as dynamically growing axes based on incoming data, are not currently possible in the \Python interfaces to \boosthistogram because a different memory model would be needed.

Until now, histograms have been stored in memory as a rectilinear $N$-dimensional array of values. Such an array cannot grow dynamically, which prevents growing axes and sparse representations of very large histograms. We discussed new memory layouts, such as \texttt{scipy.sparse} \arrays, \Blosctwo-compressed arrays, in-memory databases, and file-based formats like \HDFf (\hfivepy)~\cite{h5py} and \Zarr~\cite{zarr}. Some of these have internal search trees, some have internal chunking, and some are simply compressed in a random-accessible way to allow data to grow. These formats are particularly attractive when bin access is not uniform but concentrated in a small region of the hypercube. Disk formats like \HDFf and \Zarr have the added advantage that serializing and saving the histogram are not a separate step. With all of the data safely on disk, the process can address histograms that would be too large to fully load into available RAM.

Before \Dask and WMS engines (see above), event-processing jobs had to be followed by an accumulation job, which could be very time-consuming if the number of parallel tasks $n$ (and therefore histogram copies) was large. \Dask and WMS engines are both capable of tree-reduction, which adds histograms in $\mathcal{O}(\log n)$ steps instead of $\mathcal{O}(n)$ steps. However, accumulation is still an issue for very large histograms (or very large histogram sets) because enough memory may not be available for $n$ copies of the histogram data before accumulation. In fact, the \boosthistogram model of histograms with arbitrarily many axes encourages this: the number of histogram bins grows exponentially with the number of axes. For these cases, we must consider a single copy of the histogram with atomic operations to increment bin contents.

In addition to \ROOT's histogram presentation (plotting), many \Python packages provide visualization routines. Nearly all of these are based on \Matplotlib~\cite{matplotlib}, and two are widely used in HEP: \mplhep~\cite{Novak_mplhep_2020} and \plothist~\cite{Praz_plothist}. These two libraries were developed because \Matplotlib, by itself, has poor support for plotting pre-binned histograms, which are surprisingly uncommon outside of HEP. Although they have substantial overlap, they also have different areas of focus: \mplhep provides style templates to exactly match the fonts, line widths, and branding of the \lhc experiments, while \plothist provides more ways to compare data, such as measurements versus stacked MC simulations. Users will likely want features of both, so we discussed in depth the possibilities of merging the packages or keeping them separate but linking them with new protocols. In the end, the developers of both libraries decided to combine efforts in a single package.

As noted above, \ROOT will soon adhere to the UHI protocol, which will greatly enhance interoperability between \Python/\Matplotlib and \ROOT histogram filling and visualization.

Finally, we discussed histogram serialization. An upcoming release of \boosthistogram (the \Python package) will support serialization using a \JSON-based schema and bin contents in \HDFf, with the intention to port this format to other containers, such as \ROOT and \Zarr. Currently, many HEP users and developers serialize histograms by pickling them, which has performance and security issues mentioned above. We also agreed that histograms should carry arbitrary (\JSON-like) metadata, allowing new use cases to be developed without requiring all histogram libraries to support them as built-in features.

\section{RDataFrame, Coffea, and Dask}

Distributed computing has always played a crucial role in HEP, from batch systems with manual job submission to automated WLCG gridware and orchestration systems like \Dask~\cite{fineHEP, protoROOT}, \ApacheSpark~\cite{apache}, and \StellarHPX~\cite{taskscheduler1, taskscheduler2}, as well as WMSs like \Luigi and \Snakemake. Recently, orchestration systems have changed how we think about distributed computing by removing much of the boilerplate and setup required to parallelize applications. In particular, \Dask separates the process of describing computation DAGs from the process of executing them, which allows for alternative schedulers, such as \Ray~\cite{ray} and \TaskVine~\cite{taskvine}.

\Dask has gradually increased in prominence in the \Python software stack, evolving from one choice of backend (the other usually being \Spark) to a system that users address directly. Some tools have started to rely on it, as well as computing infrastructures~\cite{swan, coffeacasa, infncms} that feature integration between a \Jupyter front-end and \Dask diagnostics.

One question that we discussed at PyHEP.dev, and that deserves discussion beyond the workshop, is whether our community is best served by the default \Dask scheduler or whether we should investigate different scheduling systems like \Ray and \TaskVine. On the one hand, the \Dask scheduler has the clear advantage of being able to connect to many types of resource managers, most notably those batch systems that are already widely employed in our field, such as \HTCondor and \Slurm. This is handled through concrete implementations of cluster connection classes that all refer back to a single user-facing entry point, the \texttt{dask.distributed.Client}. This allows the entire set-up configuration to be abstracted away from users by providing cluster configuration options upfront by the computing facility, in a way that can be reused in any \Python script. On the other hand, the default \Dask scheduler orchestrates all tasks through a single-threaded process and it typically isn't used for HEP-scale, multi-user, persistent \Dask clusters; more often, scientists launch a single-user \Dask cluster for a specific project, then shut down the entire cluster. In addition to implementing a distributed scheduler, \TaskVine is being developed by computer scientists at the University of Notre Dame, with whom the HEP community is actively engaging.

The architecture of \Dask clearly separates DAG building from execution. In fact, these are two \Python packages that can be installed separately. The \Dask graph is a format that is widely used in the sciences, and can be considered a serialization format for DAGs that can be inspected and optimized by other libraries. In our discussions, however, we concluded that we do not know if the \Dask DAG format is sufficiently generic for all future scheduling systems: as a data structure, it is by nature static, which could prevent a future analysis tool form dynamically adapting to specific requirements of analyses (e.g.,\ treating different data samples) or optimizing the application (e.g.,\ with a dynamic tree-reduction step that picks up already finished tasks instead of waiting for pre-planned tasks to finish). These problems could be solved by leveraging the \Dask client and injecting custom workflows in the nodes of the graph, a technique already used by \RDataFrame, although this goes beyond the boundaries of what is directly offered by the tool.

Moreover, user experiences with \Dask in \Python HEP analyses encounter performance limitations in very large graphs that are slow to build or optimize, as well as memory issues that cause \Dask workers to fail and restart. HEP use-cases are pushing the complexity limits of \Dask DAGs, particularly when all systematic variations are included in the computation. \Dask DAGs are, after all, \Python dictionaries --- not a packed binary format. One solution that we discussed was to add an opt-in function to tell \Dask that a particular function can be computed on all partitions independently, similar to the \Coffea \texttt{Processor} before \daskawkward, so that \Dask sees a complex function with many operations as a single DAG node.

User-friendly analysis frameworks like \PocketCoffea simplify the set-up of HEP workflows by expressing them in a declarative configuration. These frameworks build on top of \Coffea and \RDataFrame to provide further abstractions for the most common workflows. Finding a good balance between user-friendly abstractions and flexibility is crucial to avoid restrictive setups.

\section{Future of PyHEP.dev}

Next year (2025), PyHEP.dev will be in the United States. In our final session, we discussed whether it should return to Princeton as in 2023, be co-located with the international SciPy Conference~\cite{SciPy} (in Tacoma, Washington, USA for 2025), or whether we should consider another venue. There were two arguments for co-locating with SciPy: it encourages PyHEP developers to interact with the larger scientific \Python community, and an increasing number of high-energy physicists have been attending SciPy in recent years. Additionally, stringent travel budgets and environmental considerations related to air travel mean that combining PyHEP and SciPy into a single trip is advantageous.

An argument against co-locating with SciPy is that the combined conference would span more than a week, which may be longer than some participants can justify. (Notably, no one was interested in making PyHEP.dev merely a Birds-of-a-Feather session within SciPy, reducing a multi-day workshop to a one hour session.) Historically, the third general PyHEP workshop was planned to be co-located with SciPy in 2020, but this was disrupted by the COVID-19 pandemic. In a sense, co-locating PyHEP.dev with SciPy in 2025 revisits that idea.

Looking further ahead, we considered future PyHEP.dev workshops in Europe, such as 2026. Connecting with ErUM-Data Hub~\cite{ErUM-Data-Hub}, our host and co-organizer this year, was incredibly fruitful, but we acknowledge that the ErUM-Data Hub team's generosity in hosting the 2024 workshop was a one-time event, as ErUM-Data Hub must diversify across scientific domains. We also considered the possibility of co-locating with EuroSciPy~\cite{EuroSciPy}, though this remains very provisional.

Another idea raised was to consider locations beyond the United States and Europe, to enable broader participation from collaborators. With the success of HSF-India~\cite{hsfindia} and strong interest from the Indian community in computation in HEP, we discussed the possibility of a future PyHEP.dev in India. However, this must be balanced against the travel constraints, budgets, and CO$_2$ emissions associated with the core PyHEP developers traveling longer distances.

\section{Summary}

The workshop was a great success. The format of short morning talks, longer afternoon discussions, and hackathon free time led to many issues being addressed, and some even resolved through code.

\section{Acknowledgments}

We heartily thank ErUM-Data Hub for their gracious assistance in locally organizing and financially supporting the workshop. It couldn't have been done without all of your help!

Also, we thank the Python Software Foundation (PSF) and the National Science Foundation (grants OAC-2103945 and PHY-2323298) for their financial support.

\sloppy
\raggedright
\clearpage
\printbibliography[title={References},heading=bibintoc]

\end{document}